\documentclass[11pt, a4paper]{article}
\usepackage{authblk}
\usepackage{xcolor}
\usepackage{graphicx}
\usepackage[a4paper, top=0.5in, bottom=0.75in, left=0.75in, right=0.75in]{geometry}
\usepackage{placeins}
\usepackage{amssymb}
\usepackage{tikz-cd}
\usepackage{tikz-3dplot}
\usetikzlibrary{arrows.meta, decorations.pathmorphing}
\usepackage{amsmath}
\usepackage{stmaryrd}
\usepackage{amsthm}
\usepackage{enumitem}

\usepackage{xcolor}

\usepackage{accents}
\usepackage{amssymb}
\usepackage{bm}
\usepackage{sidenotes}
\usepackage{url}

\usepackage{hyperref}
\hypersetup{
  colorlinks = true,
  linkcolor=darkblue, citecolor=darkblue,
  urlcolor=black
}

\definecolor{darkblue}{rgb}{0.05,0.25,0.65}
\definecolor{darkgreen}{RGB}{20,140,10}
\definecolor{lightgray}{rgb}{0.9,0.9,0.9}
\definecolor{darkorange}{RGB}{200,100,5}
\definecolor{darkyellow}{rgb}{.91,.91,0}
\definecolor{lightolive}{RGB}{225, 220, 185}

\newlength{\dhatheight}

\newtheorem{theorem}{Theorem}[section]

\theoremstyle{definition}

\newtheorem{remark}[theorem]{Remark}

\newcommand{\dd}{\mathrm{d}}

\newcommand{\Hodge}{\star_5}

\begin{document}
\title{M5 brane to D4 brane via cyclification of rational relative 3-cohomotopy.}

\author{Pinak Banerjee\thanks{pinakb24@vt.edu}}
\affil{Department of Physics,
\\
Virginia Tech,
\\
850 West Campus Drive,  
\\
Blacksburg,
VA 24061, USA}
\maketitle
\begin{abstract}
In this article, we start by re-deriving the equations of motion and Bianchi identities for the abelian D4 brane worldvolume. Noting that the 4-flux in M-Theory is rationally flux quantized in a non-abelian cohomology theory called 4-cohomotopy, and the three-flux on the M5 brane worldvolume in rational (twisted) 3-cohomotopy, we compute the minimal model for the cyclification of the quaternionic Hopf fibration which encodes the Bianchi identities for the fluxes on the D4 brane worldvolume after double dimensional reduction. The two pictures can be mapped to each other, and thus at the rational level, we conjecture a non-abelian relative cohomology theory for the D4 brane, fibered over the 10d Type IIA spacetime fluxes.
\end{abstract}
\section{Introduction.}
Flux quantization in physics is very much essential for understanding the field content of the theory on higher intersections in order to make the theory globally well-defined\cite{SS25-Flux},\cite{A-85}. That is why we are interested in understanding the concept of non-abelian cohomology\cite[\S 2.1]{FSS23-Char}, which seems to capture the flux quantization laws and the Bianchi identities of the theory. This can shed some light on non-perturbative phenomena, and it can go a long way towards understanding of the strongly-coupled systems.

Now, the quantization law for M theory\cite{W-95,U,4,Duff96,Duff99,BST,MS-05} has been a long-standing question, and has received little attention in the literature to the best of our knowledge (see \cite{Wi-Flux} for earlier attempts). Finally, it has been observed that the 11d spacetime M theory Bianchi identities $\dd G_4=0, \dd G_7=-\tfrac12 G_4\wedge G_4$ resemble the minimal model of the rational homotopy type of the 4-sphere. The lift to integral full non-abelian cohomology theory for M theory charges has been conjectured to be (unstable) 4-Co-homotopy. It is defined as the homotopy classes of maps from the spacetime to 4-sphere $\pi^4(\mathrm{X}):=\pi_0 \mathrm{Map(X, S^4)}$\cite[\S 2.10]{FSS23-Char}. This is usually called {\it Hypothesis H}\cite[\S 2.5]{Sa-13},\cite{GSS24-SuGra}. By coupling to background gravity, we do get interesting results from twisted cohomotopy\cite[\S 2.41]{FSS23-Char} like anomaly cancellation on 8-manifolds\cite{FSS20-H}.

Now, it is well known that the M theory on a circle gives rise to Type IIA String theory. From the rational 4-cohomotopy perspective, there is a minimal model associated with the cyclic loop spaces $\mathrm{LS^4\sslash S^1}$, or in short cyclification $\mathrm{Cyc(X)}:= \mathrm{LX\sslash S^1}$\cite[\S 3.2]{FSS2016}, where LX denotes the free loop space of X, which produces the Bianchi identities for the spacetime RR fluxes, twisted by the NS-NS three-form flux $H_3$. This can be seen as the Chevalley-Eilenberg algebra of $H_3$ twisted K theory\cite[\S 2,3,4]{FSS2016}.

As a next step, we want to understand the M5 brane\cite{Guven,Wi97,FKPZ,PST,Berman} in M theory as a system fibered over the spacetime M theory charges, or in better words, a {\it relative} non-abelian cohomology twisted by the 11d M theory spacetime charges. The sourced Bianchi identity on the M5 brane worldvolume reads $\dd H_3=\phi^{\ast}G_4$ tells us that at least at the rational level, it is the minimal model of the 7-sphere {\it relative} to the 4-sphere. The integral lift of this M5 brane story can be taken to be in the quaternionic Hopf fibration $h_{\mathbb{H}}:S^7\rightarrow S^4$ with homotopy fiber being $ S^3$. Thus the M5 brane is said to be quantized in relative 3-cohomotopy twisted by the background spacetime 4-cohomotopy charges $(c_3,c_6)$ denoted by $\pi^{3+\phi^{\ast}(c_3,c_6)}$, where $\phi:\Sigma\rightarrow X$ is the usual embedding map throughout our article where $\Sigma$ is the worldvolume and $X$ is the 11D M theory target spacetime\cite[eq.22]{GSS24-FluxOnM5}. Twisted cohomotopy also implies (twisted) String structures on the M5 brane worldvolume\cite{FSSString}, anomaly cancellation\cite{M5Anomaly} and the WZ term\cite{FSS21-Hopf}. Also, the super exceptional embedding of the M5 brane and relation to 5d Super Yang Mills have been discussed in \cite{FSS19}. For this paper, the rational picture of the quaternionic Hopf fibration will suffice.

Now, we want to see this for the D4 brane, fibered over the 10d spacetime Type IIA fluxes. We want to see this from the {\it relative} cohomotopy point of view which encodes the Bianchi identities for the M5 brane. Upon circle reduction, the M5 brane is expected to give rise to the D4 brane\cite{D41,D42, D42'',D42', D43}. From the previous discussion, it is expected that the Bianchi identities on the D4 brane should be fibered over the spacetime Type IIA fluxes. We would expect the CE algebra of cyclification of (relative) 3-cohomotopy to produce this system of D4 brane upon double dimension reduction of the M5 brane in 11d M theory. For the relationship between cyclification and double-dimensional reduction, see \cite[\S 3]{D4C}.

We also try to see the sourced Bianchi identities for the D4 brane from it's worldvolume action, both the DBI action and the Chern-Simons action \cite{D44,D45,D46,D47}. But since the DBI action has a square root term, it is pretty cumbersome to deal with, and non-linear Bianchi identities do arise for the worldvolume field strength. For small field limit, it reduces to the usual sourced Maxwell's equations. These Bianchi identities obtained from the worldvolume action can be mapped to the equations obtained from the minimal model of the cyclification of relative 3-cohomotopy.

The structure of the paper is as follows. In section 2, we rederive the equations of motion\footnote{We will interchangeably use the names Bianchi identity and the equation of motion.} from the D4 brane abelian worldvolume actions. In section 3, we compute the Chevalley-Eilenberg algebra of the cyclification of $S^7$ relative to $S^4$, and we map the Bianchi identities to the ones derived from the worldvolume action. These two precisely can be mapped to each other, which motivates us to propose a rational relative cohomology for the system of D4 brane, fibered over the Type IIA String theory Bianchi identities, which we have takn to be rationally in $\mathrm{LS^4\sslash S^1}$, not in topological K theory twisted by the NS-NS three form flux. In section 4, we have made some concluding remarks.

\section{D4 brane worldvolume theory.}

\subsection{M5 brane on a Circle}

Let the M5 worldvolume be a product $\Sigma_6 \;=\; \Sigma_5 \times S^1$ where
y is the circle coordinate and $\eta$ be the unit 1-form corresponding to the circle.
In what follows below, we will assume $y$-independence for the fields. 

We would now split the worldvolume three-form $H_3$ on the M5 brane worldvolume along the circle direction and similarly for the bulk $G_4$ flux as
\begin{equation}
H_3 \;=\; H^{(3)} \;+\; \mathcal{F} \wedge \eta,
\qquad
G_4 \;=\; {G}^{(4)} \;+\; {G}^{(3)} \wedge \eta,
\end{equation}
with $H^{(3)} \in \Omega^3(\Sigma_5)$, $\mathcal{F} \in \Omega^2(\Sigma_5)$, $\phi^{\ast}{G}^{(4)} \in \Omega^4(\Sigma_5)$, and $\phi^{\ast}{G}^{(3)} \in \Omega^3(\Sigma_5)$.

\paragraph{Bianchi reduction.}
We know the Bianchi on the M5 worldvolume to be $\dd H_3 = \phi^{\ast}G_4$, and thus we obtain the 5D sourced equations
\begin{equation}\label{eq:5DBianchi}
{ \; \dd H^{(3)} \;=\; \phi^{\ast}{G}^{(4)},\hspace{0.25cm} \dd \mathcal{F} \;=\; \phi^{\ast}{G}^{(3)}. \;}
\end{equation}

We now fix the conventions of the Hodge dual with respect to the induced metric on the worldvolume
with $\mathrm{vol}_6=\mathrm{vol}_5\wedge\eta$. We have
\[
\ast_6(H^{(3)}) = (\ast_5 H^{(3)}) \wedge \eta,
\qquad
\ast_6(\mathcal{F}\wedge\eta) = \ast_5 \mathcal{F} .
\]

Upon KK reduction to type IIA, we identify $\phi^{*}{G}^{(3)}$ with the pulled-back NSNS $3$-form $\phi^{*}H_{\mathrm{NS}}$ to the D4 brane worldvolume.
We now define the combination of the worldvolume field strength and the NS-NS two form potential pulled back to the D4 worldvolume as the gauge-invariant combination
\begin{equation}
{ \; \mathcal{F} \;:=\; f_2 + \phi^{*}B \;,}
\end{equation}
where $\phi^{*}B$ is the pulled-back NSNS $2$-form to the D4 brane and $f_2=\dd A, \dd f_2=0$. Then
\begin{equation}
\dd\mathcal{F} \;=\;\dd f_2 + \dd(\phi^{*}B) \;=\; \phi^{*}H_{\mathrm{NS}}.
\end{equation}

\subsection{Worldvolume action of the D4.}
\paragraph{DBI action.}
We will start with the abelian DBI lagrangian for the D4 worldvolume here. See \cite{D44,D45,D46,D47} for more details.

Let $g$ be the induced 5D metric on $\Sigma_5$ (which we will take to be the flat metric here), and consider the matrix
\[
M \;:=\; -g - \mathcal{F}.
\]
Here, $\mathcal{F}:= f_2 + \phi^{\ast}B_2 ,\qquad \dd \mathcal{F}= \phi^{\ast} H_{\mathrm{NS}} , \qquad \dd B_2 = H_{\mathrm NS}$
where $H_{NS}$ be the NS-NS three-form flux in Type IIA string theory and $f_2$ denotes the abelian field strength on the D4 brane worldvolume with $\dd f_2=0$.

We will set $g=\eta$ for now, as mentioned.

The abelian DBI Lagrangian density \cite[eqns. 8, 60]{D42'} on the single D4 brane worldvolume reads
\begin{equation}
\mathcal{L}_{\mathrm{DBI}}
\;=\;
-\,T_4\,\sqrt{\det(M)}.
\end{equation}
Here, $T_4$ denotes the tension of the D4 brane.

\paragraph{Variation of the determinant.}

We would like to see the equations of motion coming out of this lagrangian by varying the worldvolume gauge field A.

Varying $A$ gives $\delta \mathcal{F} = d(\delta A)$ and
\[
\delta \sqrt{\det M} = \tfrac12 \sqrt{\det M}\,\mathrm{Tr}\!\big(M^{-1}\delta M\big),
\quad \delta M = -\delta\mathcal{F}.
\]
Note that only the antisymmetric part of $M^{-1}$ contributes here. We define for our convenience
\begin{equation}
{
\ \mathcal{G}^{\,ab} \;:= 
-\sqrt{\det(M)}\;\big(M^{-1}\big)^{[ab]}, \qquad
\mathrm{where}\hspace{0.5cm} X^{[ab]} := \tfrac12\,(X^{ab}-X^{ba}) .
}
\end{equation}
Equivalently, in differential-form language this reads,
\begin{equation}
\mathcal{G} \;:=\; \tfrac12\,\mathcal{G}_{ab}\,dx^a\wedge dx^b,
\qquad
 (\ast_5 \mathcal{G})_{abc}= \frac{1}{2} \epsilon_{abcde}\mathcal{G}^{de} .
\end{equation}
where the symbol $\ast_5$ denotes the Hodge dual with respect to the 5d worldvolume metric, and the indices a,b,c.. denote the worldvolume indices.

\paragraph{Equations of motion.}

We now want to figure out the equations of the worldvolume gauge field from the action
\begin{equation}
    \begin{array}{l}
    \delta S = \int \delta \mathcal{L}_{\mathrm{DBI}} = -\frac{T_4}{2}\int \sqrt{\det(M)}\;\big(M^{-1}\big)^{ab}\delta M _{ab}.
    \end{array}
\end{equation}
Since $$\delta M _{ab}= \delta \mathcal{F}_{ab}$$
we need to consider only the antisymmetric part of the matrix $M$.

So we get the variation of the action to be
\begin{equation}
    \begin{array}{l}
    \delta S = \int \delta \mathcal{L}_{\mathrm{DBI}} = \frac{T_4}{2}\int \mathcal{G}^{ab}\delta \mathcal{F} _{ab}.
    \end{array}
\end{equation}
Now, we integrate by parts the variation  $\delta\mathcal{F}=d(\delta A)$ in $\delta S=  {T_4}\int \mathcal{G}^{ab}\partial_{a}\delta A_{b}$, and assuming that the worldvolume gauge field $A$ vanishes at the boundary of the worldvolume one gets
$$\delta S=  -{T_4}\int \partial_{a}\mathcal{G}^{ab}\delta A_{b}$$ or equivalently in differential form language
$$\delta S=  -{T_4}\int \dd (\ast_{5}\mathcal{G})\wedge \delta A.$$
Thus for generic variation $\delta A$, we obtain the equations of motion
\begin{equation}\label{DBIeom}
{
\ \partial_a \mathcal{G}^{\,ab} = 0
\quad\Longleftrightarrow\quad
\dd\big(*_5 \mathcal{G}\big) = 0 .
}
\end{equation}
On top of this, we have the usual Bianchi identity to be
\begin{equation}
{
    \dd \mathcal{F}= \phi^{\ast}H_{\mathrm{NS}}.
    }
\end{equation}
 For the small field limit, the DBI theory reduces to the usual Maxwell theory.

\paragraph{Series expansion for the DBI action.}
Let's define $X:=\eta^{-1} \mathcal F$ . Then we have
\[
\sqrt{\det(-\eta-\mathcal F)}=\sqrt{-\eta}\;\sqrt{\det( 1+X)}
=\;\exp\!\Big(\tfrac12\mathrm{Tr}\ln( 1+X)\Big).
\]
Since the trace of all odd powers of the antisymmetric matrix vanishes $\mathrm{Tr}(\mathcal{F}^{2k+1})=0$, we get
\begin{equation}
\label{eq:sqrtseries}
\sqrt{\det (1+X)}=1-\tfrac14\mathrm{Tr}(X^2)-\tfrac18\mathrm{Tr}(X^4)+\tfrac1{32}\big(\mathrm{Tr}X^2\big)^2+\mathcal O(X^6).
\end{equation}
Also the antisymmetric part of $M$ is, 
\begin{equation}
\label{eq:MinvAntiSeries}
\big(M^{-1}\big)^{[ab]}=\big(-X-X^3-X^5+\mathcal O(X^7)\big)^{ab}.
\end{equation}
Multiplying \eqref{eq:sqrtseries} and \eqref{eq:MinvAntiSeries} gives, with all indices lowered,
\begin{align}
\label{eq:DBIseries}
\mathcal G_{ab}
&=\;\Big[
\ \mathcal F_{ab}
+(\mathcal F^3)_{ab}
+(\mathcal F^5)_{ab}
-\tfrac14(\mathrm{tr}\,\mathcal F^2)\,\mathcal F_{ab}
-\tfrac14(\mathrm{tr}\,\mathcal F^2)\,(\mathcal F^3)_{ab}
\nonumber\\[-2pt]
&\hspace{7.6em}
-\tfrac18(\mathrm{tr}\,\mathcal F^4)\,\mathcal F_{ab}
+\tfrac1{32}(\mathrm{tr}\,\mathcal F^2)^2\,\mathcal F_{ab}
+\mathcal O(\mathcal F^7)\Big],
\end{align}
where $(\mathcal F^3)_{ab}:=\mathcal F_{a}{}^{c}\mathcal F_{c}{}^{d}\mathcal F_{db}$,
$\mathrm{tr}\,\mathcal F^2:=\mathcal F_{ab}\mathcal F^{ba}$, and similarly for $\mathrm{tr}\,\mathcal F^4$.

\begin{align}
\label{eq:DBIseries1}
\ast_{5}(\mathcal G_{ab})
&=\;\Big[
\ \ast_{5}(\mathcal F_{ab})
+\ast_{5}(\mathcal F^3_{ab})
-\tfrac14 \Hodge(\mathrm{tr}\,(\mathcal F^2)\,\mathcal F_{ab})
+\mathcal O(\mathcal F^5)\Big],
\end{align}
Now from \eqref{DBIeom}, we would naively expect
\begin{align}
\label{eq:DBIeom}
0 = \dd \ast_{5}(\mathcal G_{ab})
&=\dd \Big[
\ \ast_{5}(\mathcal F_{ab})
+\ast_{5}(\mathcal F^3_{ab})
-\tfrac14 \Hodge(\mathrm{tr}\,(\mathcal F^2)\,\mathcal F_{ab})
+\mathcal O(\mathcal F^5)\Big].
\end{align}
which is not true.

\begin{remark}{In the dictionary, if we are to relate $\ast_{5}\mathcal{G}$ and $H^{(3)}$, we will find one inconsistency: $\\d \ast_{5}\mathcal{G}=0$ while $\dd H^{(3)}\neq 0$. So there need to be some extra terms on the right-hand side of $\dd \ast_{5}\mathcal{G}$ already not captured by the DBI action, and this tells us that the DBI action is not enough. We need to add the CS action for the D4 brane worldvolume as well.}
\end{remark}

\paragraph{CS action for the D4 brane worldvolume.}

We now write out the CS action \cite[eq. 60]{D42'} for the D4 brane worldvolume here

\begin{equation}
    S_{\mathrm{CS}}= -T_4\int \sum_{q=0}^{5}\phi^{\ast}C_{q}\wedge e^{\mathcal{F}}= -T_4\int \phi^{\ast}C_{5}+ \phi^{\ast}C_{3}\wedge\mathcal{F}+ \frac{1}{2}\phi^{\ast}C_{1}\wedge\mathcal{F}\wedge \mathcal{F}.
    \end{equation}
    We need to vary the worldvolume gauge field to find the variation of the CS action, and this gives
    \begin{equation}
      \delta S_{\mathrm{CS}}= -T_4\int  \phi^{\ast}C_{3}\wedge \delta\mathcal{F}+ \phi^{\ast}C_{1}\wedge\mathcal{F}\wedge \delta\mathcal{F}  
    \end{equation}
    Using $\delta\mathcal{F}= \dd (\delta A)$ and implementing again the usual integration by parts, we obtain
    \begin{equation}
    \begin{array}{l}
      \delta S_{\mathrm{CS}}= -T_4\int  \dd (\phi^{*}C_{3})\wedge \delta{A}+ \dd (\phi^{\ast}C_{1}\wedge\mathcal{F})\wedge \delta{A} \\
      = -T_4 \int (\phi^{\ast} F_4+ \phi^{\ast} F_2 \wedge \mathcal{F})\wedge \delta A
      \end{array}
    \end{equation}
    where $F_4, F_2$ denote the spacetime RR fluxes corresponding to the RR potentials $C_3, C_1$ in Type IIA String theory, and $F_4=dC_3- C_1\wedge H_3$.
\paragraph{Full variation of the actions.}
    Now, taking the combined variation of both the DBI and CS actions into account, we obtain from $\delta S_{\mathrm{DBI}}+ \delta S_{\mathrm{CS}}=0$ the following equation for the D4 worldvolume:
    \begin{equation}
    \label{totalvariation}
    {
        \dd (\ast_{5} \mathcal{G})=  -(\phi^{\ast}F_4+\phi^{\ast}F_2\wedge \mathcal{F}).
        }
    \end{equation}
    We can express the full equation in terms of the worldvolume field strengths below following \eqref{eq:DBIeom}: 
\begin{equation}
\label{eq:DBIeomfinal}
{
\phi^{\ast}F_4+\phi^{\ast}F_2\wedge \mathcal{F}
=-\dd \Big[
\ \ast_{5}(\mathcal F_{\mu\nu})
+\ast_{5}(\mathcal F^3_{\mu\nu})
-\tfrac14 \Hodge(\mathrm{tr}\,(\mathcal F^2)\,\mathcal F_{\mu\nu})
+\mathcal O(\mathcal F^5)\Big].
}
\end{equation}
    \begin{remark}{Thus, we can now consistently identify  $\phi^{\ast}G^{(4)}$ with $(\phi^{\ast}F_4+\phi^{\ast}F_2\wedge \mathcal{F})$. This makes sense as the M theory four-form flux gives rise to the RR 4-flux in Type IIA theory when no single leg is on the compactification circle.}
    \end{remark}

  Thus, to wrap things up, on the D4 brane worldvolume we have the Bianchi identities (see also \cite[eqns. 19,20]{D41})
  \begin{equation}
  \label{NLG}
      \begin{array}{l}
      \dd \mathcal{F}=\phi^{\ast} H_{NS}\\
      \dd (\ast_{5} \mathcal{G})= - (\phi^{\ast}F_4+\phi^{\ast}F_2\wedge \mathcal{F})
      \end{array}
  \end{equation}
If we want to truncate to {\it linear} order in $\mathcal{F}$, using \eqref{eq:DBIseries} we get the Bianchi identities on the D4 brane worldvolume to be
\begin{equation}
\label{FinalBianchi}
      \begin{array}{l}
      \dd \mathcal{F}=\phi^{\ast} H_{NS}\\
      \dd (\ast_{5} \mathcal{F})= -(\phi^{\ast}F_4+\phi^{\ast}F_2\wedge \mathcal{F}).
      \end{array}
  \end{equation}

  \section{D4 brane worldvolume fluxes from cohomotopy.}

We know from Hypothesis H that the single M5 brane worldvolume theory is charge quantized in the quaternionic Hopf fibration $S^3\rightarrow S^7\rightarrow S^4$, fibered over the 11d M theory spacetime information encoded by 4-cohomotopy. 

We would like to understand the cohomology for a single D4 brane in Type IIA String Theory, from cohomotopy, at least at the rational level.

For rational homotopy theory, see \cite[\S 3.2]{FSS23-Char}.

For CE algebras of loop spaces and cyclification, see \cite[\S 3.2]{FSS2016},\cite[\S 3]{D4C}.

For M theory, we have the CE algebra\cite[\S 3.3]{FSS2016}
\begin{equation}
    \mathrm{CE}(\mathfrak{l}S^4)=(\Lambda^\bullet(g_4,g_7),\dd g_4=0, \dd g_7=-\tfrac12 g_4^2).
\end{equation}
Under the dgca morphism $\mathrm{CE}(\mathfrak{l}S^4)\rightarrow \mathrm{\Omega^\bullet_{\mathrm{dR}}(X^{1,10})}$ with $g_4\mapsto G_4, g_7\mapsto G_7$ we recover the usual Bianchi identities for 11d spacetime M theory fluxes 
$$\dd G_4=0, \hspace{0.25cm}\dd G_7=-\tfrac12 G_4^2.$$
Now, if we want to reduce it on a circle to get the Type IIA picture, we have\cite[\S 3.3]{FSS2016}
\begin{equation}
\begin{array}{l}
    \mathrm{CE}(\mathfrak{l}(\mathrm{L}S^4\sslash S^1))=\Big(\Lambda^\bullet(\omega_2,g_4,g_7,g_3,g_6; \dd'); \dd' g_4= \omega_2 g_3, \dd' g_7= -\tfrac12 g_4^2+\omega_2 g_6,\\ \dd' g_3=0, \dd'g_6= g_3 g_4, \dd' \omega_2=0 \Big)
    \end{array}
\end{equation}
where $g_3=sg_4, g_6=sg_7$.

Under the dgca morphism $\mathrm{CE}(\mathfrak{l}\mathrm{LS^4\sslash S^1})\rightarrow \mathrm{\Omega^\bullet_{\mathrm{dR}}(X^{1,9})}$ with $\omega_2\mapsto F_2, g_4\mapsto F_4, g_3\mapsto -H_3, g_6\mapsto F_6, g_7\mapsto H_7$ we recover the usual Bianchi identities for 10d spacetime Type IIA  theory fluxes 
$$\dd F_2=0, \hspace{0.25cm}\dd F_4= -F_2\wedge H_3,\hspace{0.25cm} \dd F_6= -F_4\wedge H_3,\hspace{0.25cm}\dd H_3=0, \hspace{0.25cm}\dd H_7=-\tfrac12 F_4^2+ F_2\wedge F_6.$$
Here, $F_{2,4,6}$ denote the spacetime RR fluxes and $H_3, H_7$ are the NS-NS three form flux and after duality constraint from 11d $G_7=\ast_{11} G_4$ it's Hodge dual in 10d respectively; $H_7=\ast_{10}H_3$.

These equations encode the Bianchi identities for the Type IIA spacetime RR fluxes in even degrees till degree 6, twisted by the NS-NS three-form flux. The last equation encodes the Chern-Simons term for the NS5 brane \cite[page 12]{FSS2016}\footnote{Our equation for $F_6$ differs from \cite{FSS2016} by a factor of $-\frac{1}{2}$.}.

\begin{remark}: This CE algebra is `almost' the same as $\mathrm{CE}(\mathfrak{l}(\mathrm{KU_0\sslash BU(1)}))_{\le 6}$ except the fact that the $g_7$ does not arise in the (6-truncated) twisted K theory. For this CE algebra, we have (see \cite[\S 4.10]{FSS2016}):
\begin{equation}
\begin{array}{l}
    \mathrm{CE}(\mathfrak{l}(\mathrm{KU_0\sslash BU(1)}))_{\le 6}=\Big(\Lambda^\bullet(g_2,g_4,g_6, h_3; \dd); \dd g_4= g_2 h_3, \\  \dd g_6= h_3 g_4, \dd g_2=0, \dd h_3=0 \Big).
    \end{array}
\end{equation}
Thus, we find
$$\mathrm{CE}(\mathfrak{l}(\mathrm{L}S^4\sslash S^1))/\langle g_7\rangle \cong \mathrm{CE}(\mathfrak{l}(\mathrm{KU_0\sslash BU(1)}))_{\le 6}.$$
See \cite[\S 4]{FSS2016} for further details about the Type IIA supercocycles.
\end{remark}

Now for the single M5 brane, we have the {\it relative} CE algebra (see \cite[eq. 18]{GSS24-FluxOnM5})
\begin{equation}
    \mathrm{CE}(\mathfrak{l}_{S^4}S^7)=(\Lambda^\bullet(h_3,g_4,g_7);\dd h_3=g_4, \dd g_4=0, \dd g_7= -\tfrac12 g_4^2)
\end{equation}
Under the dgca morphism $\mathrm{CE}(\mathfrak{l}_{S^4}S^7)\rightarrow \mathrm{\Omega^\bullet_{\mathrm{dR}}(\Sigma^{\mathrm{M}5};X^{1,10})}$\footnote{We do mean relative non-abelian rational cohomology here by this notation.} with $g_4\mapsto G_4, g_7\mapsto G_7, h_3\mapsto H_3$ we recover the usual Bianchi identities for the M5 brane worldvolume flux fibered over the 11d spacetime M theory fluxes\cite[eq. 132]{GSS24-FluxOnM5} 
$$\dd H_3=\phi^\ast G_4, \dd G_4=0, \hspace{0.25cm}\dd G_7=-\tfrac12 G_4^2.$$
Next we compute \begin{equation}
    \mathrm{CE}(\mathfrak{l}\mathrm{L}S^7_{/S^4})=(\Lambda^\bullet(h_3,g_4,g_7,h_2,g_3,g_6);\dd h_3=g_4, \dd g_4=0, \dd g_7= -\tfrac12 g_4^2, \dd h_2= -g_3, \dd g_3=0, \dd g_6=  g_4 g_3)
\end{equation}
where $h_2=s h_3, g_3=s g_4, g_6=s g_7$.

  {This notation $\mathrm{CE}(\mathfrak{l}\mathrm{L}S^7_{/S^4})$ means the minimal model of (the rational homotopy type of) $\mathrm{LS^7}$ {\it relative} to $\mathrm{LS^4}$ corresponding to the quaternionic Hopf fibration.}

Now, we are reducing on the M theory circle to go to Type IIA String Theory picture, we homotopy quotient by the $S^1$ action for the double dimensional reduction, and
we have the corresponding CE algebra
\begin{equation}
\label{relativeCE}
\begin{array}{l}
    \mathrm{CE}(\mathfrak{l}(\mathrm{L}S^7_{/S^4}\sslash S^1))=\Big(\Lambda^\bullet(\omega_2,h_3,g_4,g_7,h_2,g_3,g_6; \dd_{\mathrm{cyc}}); \dd_{\mathrm{cyc}} g_4= \omega_2 g_3, \dd_{\mathrm{cyc}} g_7= -\tfrac12 g_4^2+\omega_2 g_6,\\ \dd_{\mathrm{cyc}} h_3= g_4+\omega_2 h_2, \dd_{\mathrm{cyc}} g_3=0, \dd_{\mathrm{cyc}}g_6= g_3 g_4, \dd_{\mathrm{cyc}}h_2=-g_3, \dd_{\mathrm{cyc}} \omega_2=0 \Big).
    \end{array}
\end{equation}
This CE algebra is already {\it relative} to the corresponding CE algebra of the rational homotopy type of $\mathrm{LS^4\sslash S^1}$.

Under the dgca morphism $\mathrm{CE}(\mathfrak{l}\mathrm{LS^7_{/ S^4}\sslash S^1})\rightarrow \mathrm{\Omega^\bullet_{\mathrm{dR}}(\Sigma^{D4};X^{1,9})}$ with $\omega_2\mapsto F_2, g_4\mapsto F_4, g_3\mapsto -H_3, g_6\mapsto F_6, g_7\mapsto H_7, h_2\mapsto \mathcal{F}, h_3\mapsto f_3$ we recover the usual Bianchi identities for 10d spacetime Type IIA  theory fluxes 
alongside the D4 worldvolume Bianchi identities
$$\dd \mathcal{F}=\phi^{\ast}H_3, \dd f_3= \phi^{\ast}F_4+\phi^{\ast}F_2\wedge \mathcal{F}.$$
Thus we see after the duality constraint, once we take $h_2\mapsto \mathcal{F}$ and $h_3\mapsto f_3= -\ast_5 \mathcal{G}$, we recover \eqref{NLG} exactly, the Bianchi identities on the D4 brane worldvolume
$$\dd \mathcal{F}=\phi^{\ast}H_3, \dd \ast_5\mathcal{G}= -\phi^{\ast}F_4-\phi^{\ast}F_2\wedge \mathcal{F}.$$
Note this contains non-linear terms in $\mathcal{F}$.

At linear order for small fields, we recover the usual sourced Maxwell equations as in \eqref{FinalBianchi} to be
$$\dd \mathcal{F}=\phi^{\ast}H_3, \dd \ast_5\mathcal{F}= -\phi^{\ast}F_4-\phi^{\ast}F_2\wedge \mathcal{F}.$$

We have the homotopy fiber for the quaternionic Hopf fibration to be $\simeq S^3$. Here, when we turn off the background spacetime fluxes, we see the D4 brane worldvolume is rationally quantized in $\mathrm{L}S^3$ whose CE algebra is given by
\begin{equation}
    \mathrm{CE}(\mathfrak{l}\mathrm{L}S^3)= (\Lambda^\bullet(h_3,h_2);\dd h_3=0, \dd h_2=0)
\end{equation}
which is rationally $\simeq K(\mathbb{Q},2)\times K(\mathbb{Q},3)$.

The $h_2$, $h_3$ correspond to the U(1) flux on the worldvolume and it's dual (after duality constraint) in 5d; thus we recover the usual (without background fluxes) 5d SYM Bianchi identities $$\dd f_2=0, \dd f_3=0, \hspace{0.5cm}\mathrm{where\hspace{0.1cm} after \, duality\, constraint}\,\hspace{0.5cm}f_3\sim\ast_5 f_2.$$

Thus, {\it rationally } we propose the relative non-abelian cohomology for a single abelian D4 brane fibered over the Type IIA String Theory spacetime fluxes\footnote{Here, at the rational level we are taking the LS$^4\sslash S^1$ story instead of the BU(1) twisted topological K theory picture, which is more widely recognised as the cohomology theory for Type IIA. In this picture, we have shown the relative cohomology necessary for the D4 brane, at least rationally. Also, rational non-abelian cohomology and de Rham non-abelian cohomology are related to each other via deRham theorem \cite[\S 3.3]{FSS23-Char}.}

 \[
\begin{tikzcd}[column sep=125pt, row sep=125pt]
{\Sigma^{\mathrm{D4}}} \arrow[r, "\mathrm{\Big(\phi^*F_{2i,i\le 3},\phi^* H_3,\phi^* H_7,\mathcal{F},f_3\Big)}"] \arrow[d, "\phi"] & \Omega^1_{\mathrm{dR}}(-;\mathfrak{l}(\mathrm{L}S^7_{/S^4}\sslash S^1))_\mathrm{clsd}\arrow[d, "(\mathfrak{l}\mathrm{Lh_{\mathbb{H}}})_\ast"] \\
X^{1,9} \arrow[r, "\big(\mathrm{F_{2i,i\le 3},H_3,H_7}\big)"] & \Omega^1_{\mathrm{dR}}(-;\mathfrak{l}(\mathrm{LS}^4 \sslash S^1))_{\mathrm{clsd}}.
\end{tikzcd}
\]

\begin{remark}We have not mentioned anything about the {\it integral} lift of this relative cohomology story. Can we take the obvious choice of uplift from this rational picture? We are not overdemanding anything here.
\end{remark}

\section{Conclusion.}
In this brief note, we have re-derived in section 2 the equation of motion for the abelian D4 brane worldvolume fluxes in the presence of the Type IIA spacetime fluxes. In section 3, we have taken into fact that the M5 brane is quantized rationally in the quaternionic Hopf fibration, and from there we have used the corresponding CE algebra of the cyclification of the quaternionic Hopf fibration to generate the Type IIA spacetime fluxes and the Bianchi identities for the D4 brane worldvolume fluxes at the same time. The analysis exactly seems to match the one we have derived from the abelian worldvolume action for the D4 brane. Thus, we have closed the section by proposing a rational relative non-abelian cohomology for the D4 brane, fibered over the Type IIA 10d spacetime fluxes. It is worth mentioning here that we have that the rational non-abelian cohomology for the spacetime Type IIA fluxes in $\mathrm{LS^4}\sslash S^1$ , and not in $\mathrm{KU_0\sslash BU(1)}$ , and then we have fibered the D4 brane fluxes over this as a {\it relative} cohomology. It would be interesting to understand it's integral lift.

\end{document}